Three-dimensional electromagnetic metamaterials with non-Maxwellian effective fields

Jonghwa Shin, Jung-Tsung Shen, and Shanhui Fan

*Ginzton Laboratory and Department of Electrical Engineering, Stanford University, Stanford, CA 94305*

It is commonly assumed that the long-wavelength limit of a metamaterial can always be described in terms of effective permeability and permittivity tensors. This assumption holds true in all metamaterials considered up to now. Here we report that this assumption is false—there exist an entirely new class of metamaterials consisting of multiple interlocking disconnected metal networks, for which the effective long-wavelength theory is local, but the effective field is non-Maxwellian, and possesses much more internal degrees of freedom than effective Maxwellian fields in a homogeneous medium.

Metamaterials, which consist of a periodic array of metallic or dielectric elements, exhibit important electromagnetic effects [1], especially in the limit where the wavelength of radiation is much larger than the periodicity. In such limit one generally illustrates the physics of metamaterials by constructing an equivalent *homogenous* medium. It is commonly believed that such equivalent homogenous medium can always be described by Maxwell equations with an effective permittivity tensor $\varepsilon$ and permeability tensor $\mu$ [1,2]. Consequently the corresponding effective field for any metamaterial is Maxwellian. This belief is consistent with the studies of homogenization [2,3] since the work of Maxwell-Garnett a hundred years ago, as well as with all recent researches, which have shown that the effective $\varepsilon$ and $\mu$ tensors can exhibit a remarkably rich set of local or nonlocal behaviors [4-10], in spite of several subtleties regarding homogenization for certain metallic systems [11,12].

In this Letter we report, however, that this assumption is false—there in fact exist an entirely new class of metamaterials for which the effective long-wavelength theory is distinctively non-Maxwellian. For these structures the effective fields feature large numbers of internal degrees of freedom. These new structures point to previously unexplored regimes of optical physics and device applications. Moreover, in connection with the modern developments of electronic theory in condensed matter physics, in which the symmetry of the low-energy (i.e., long-wavelength) excitation can be substantially different from the underlying microscopic Hamiltonian [13], this work suggests that similar developments can be fruitful in electromagnetic theory, in which new effective theory emerges from topological complexity that is commonly occurring in metamaterials.

To establish our results, we first note that describing the long-wavelength properties of a metamaterial in terms of an effective Maxwellian field imposes general constraints on its allowed electromagnetic behaviors. For example, if a structure has symmetry of the cubic ($O_h$) or pyritohedral ($T_h$) point group [14], one can rigorously prove that $\varepsilon$ and $\mu$ must be scalar [15,16]. The effective field supports either zero [17,18] or two [19] propagating modes, depending on the sign of the product $\varepsilon\mu$. The bands are degenerate and isotropic, i.e., the iso-frequency surface in k-space is a single sphere.

Here we study three-dimensional structures consisting of multiple interlocking, disconnected metal networks. Each network by itself is connected in full three dimensions. (Fig. 1) These structures all possess a cubic unit cell and either the cubic or pyritohedral point group symmetry. Each structure is specifically chosen to highlight at least one remarkable feature that cannot be accounted for by a Maxwellian effective field.

In particular, there is no constraint on the number of propagating modes (Fig.1a-c). Also, in spite of the structural symmetry, these modes can be either non-degenerate (Fig.1b), or anisotropic (Fig. 1c).

In order to describe this new class of metamaterials, we construct a local effective field theory based on a quasi-static approach [9,20], which is exact in the long-wavelength limit. The electro- and magneto-static properties of our $N$-network systems can be described as $\mathbf{Q}=\mathbf{CV}$ and $\mathbf{A}=\mathbf{LI}$, i.e.,

$$Q_i = \sum_j C_{ij} V_j \text{ and } A_{jv} = \sum_{iu} L_{jv,iu} I_{iu}, \qquad (1)$$

respectively $\left(i, j \in \{1, \ldots, N\}; u, v \in \{x, y, z\}\right)$. Here, the subscript $i$ or $j$ labels an individual network. $V_i$ is the voltage. $Q_i$ is the total charge in a unit cell divided by the unit cell volume. $I_{iu}$ is the total current through a unit cell surface normal to the $u$-direction, divided by the area of the surface. Following the concept of inductance, the magnetic potential is defined as an average of the vector potential on an integration path along the conductor:

$$A_{jv} = a^{-1} \int_{\vec{r}_j}^{\vec{r}_j + a\vec{v}} \vec{A} \cdot d\vec{l}, \qquad (2)$$

where $a$ is the unit cell length and the integration path remains on the $j$-th network. Both $\mathbf{C}$ and $\mathbf{L}$ are real and symmetric due to reciprocity. [$\mathbf{L}$ is represented as a $3N$-by-$3N$ matrix with $\mathbf{A} \equiv \left(A_{1x}\ A_{1y}\ \cdots\ A_{Nz}\right)^{\mathrm{T}}$ and $\mathbf{I} \equiv \left(I_{1x}\ I_{1y}\ \cdots\ I_{Nz}\right)^{\mathrm{T}}$, where the superscript T denotes transpose.]

We now describe the properties of electromagnetic waves. Since the microscopic electric field vanishes inside the conductor, $\vec{E} = -\nabla V - \partial \vec{A}/\partial t = 0$. Integrating this relation as in Eq. (2), we obtain

$$\nabla \mathbf{V} = -\frac{\partial \mathbf{A}}{\partial t}. \tag{3}$$

Eq. (3) is valid for any integration path for $\vec{A}$ along the conductors. Also, the microscopic current density and charge density are related by $\nabla \cdot \vec{J} = -\partial \rho / \partial t$. Again, by integration, we have

$$\nabla \cdot \mathbf{I} = -\partial \mathbf{Q}/\partial t. \tag{4}$$

Both Eqs. (3) and (4) are exact for all frequencies. They are not restricted to the quasi-static limit.

In the quasi-static limit, the dynamic variables $\mathbf{V}$, $\mathbf{Q}$, $\mathbf{I}$, and $\mathbf{A}$ vary at a length scale that is much longer compared with the periodicity. Thus, within each unit cell, Eq. (1) remains valid. Combining this with Eqs. (3) and (4), we arrive at the equation for $\mathbf{V}$ as

$$\frac{\partial^2 \mathbf{V}}{\partial t^2} = \mathbf{C}^{-1} \nabla \cdot \left( \mathbf{L}^{-1} \nabla \mathbf{V} \right). \tag{5}$$

In deriving Eq. (5), there is an important subtlety regarding the matrices $\mathbf{C}$ and $\mathbf{L}$. The capacitance matrix $\mathbf{C}$ has a zero eigenvalue, since a constant shift in the voltages of all networks does not contribute to the difference in charge distributions between conductors. Also, three of the eigenvalues of the inductance matrix $\mathbf{L}$ is infinite, since a constant current in a fixed direction results in infinite vector potential. However, wave solutions have zero total voltage and current in each unit cell, i.e., $\sum_i V_i = 0, \sum_i \vec{I}_i = 0$.

Thus, within the subspace of wave solutions, the inverses of matrices **C** and **L** are appropriately defined after projecting out the eigenvectors associated with the zero or infinite eigenvalues. The inverses in Eq. (5) are defined in this way. To emphasize this, below we refer to $\mathbf{C}^{-1}$ and $\mathbf{L}^{-1}$ as **S** and **U**, respectively. The eigenvalues of **S** and **U** are all finite.

Equation (5) describes the dynamics of an effective field **V** that possesses $N-1$ independent components, in an effective uniform medium characterized by the matrices **S** and **U**. In order to obtain the dispersion relations, one assumes that the voltage varies as $e^{j(\omega t - \vec{k}\cdot\vec{r})}$. The frequencies of the modes at a given $\vec{k}$ can then be determined by solving the corresponding matrix eigenvalue problem defined through Eq. (5). The matrices **S** and **U**, which describe the electric and magnetic response of the structure, play analogous roles to the inverses of $\varepsilon$ and $\mu$ tensors in the Maxwell equations. They are constant matrices with no wavevector dependency. Hence, we refer to the field theory defined by Eq. (5) as local.

The components of **V** represent the internal degrees of freedom of the effective fields. They are analogous to polarization for Maxwellian fields. However, unlike polarizations, these internal degrees of freedom are designable by geometry, and hence can exhibit much richer behaviors.

To highlight the unique electromagnetic properties Eq. (5) describes, we calculate the electromagnetic bands of the structures shown in Fig. 1 using finite difference time domain (FDTD) methods as well, which solve the underlying Maxwell equations from first principles without any uncontrolled approximation. The computational domain corresponds to one cubic unit cell with Bloch-periodic boundary conditions [19], discretized with 48 grid points in each direction. The simulations yield a band structure assuming a cubic unit cell. Since the cubic unit cell contains all networks, the results thus obtained can be directly compared to the effective field theory.

*Case (1): Two-network crystal.* The structure consists of two metallic networks and has a point group of a cube ($O_h$)[Fig. 1(a)]. Each network consists of wires along the edges of a cube. The two networks are displaced from one another by $\sqrt{3}a/2$ along the $[111]$ direction. With two conductors, the effective field has only one degree of freedom and the **S** matrix has a simple form of $S_{12}\begin{pmatrix}-1 & 1\\ 1 & -1\end{pmatrix}$. Since each network has fourfold rotational symmetry axes along $x$, $y$, and $z$ directions that pass through the origin, one can prove that $U_{iu,jv} = U_{ij}\delta_{uv}$. Therefore, our theory [Eq. (5)] predicts the structure has one low-frequency mode with an isotropic dispersion relation, in consistency with FDTD simulations [Fig. 2(a)]. As a further proof of the single-mode nature, we plot the electric field profile at a representative k-point in Fig. 2(b). The field indeed has the full symmetry of the lattice.

The two-network system thus supports an isotropic scalar effective field. In using optical waves for communication, polarization dependency due to the vector nature of electromagnetic fields is a major detriment. In connection with the important attempts of creating scalar field in one dimension using novel fiber structures [21], our work suggests a route towards completely erasing polarization dependency in all three dimensions. In particular, the field profile in a unit cell closely matches that of those novel fiber structures.

*Case (2): N-network cubic crystal.* The structure consists of $N$ metallic networks, where $N$ can be arbitrarily large [Fig. 1(b)]. Each network has metal wires along $x$, $y$, and $z$ directions. For the $i$-th network, these wires meet at $(\pm x_i, \pm x_i, \pm x_i)$ in each cubic unit cell. Having $x_i$'s differ more than the thickness of the wires ensures that networks are not touching. Such a structure supports $N-1$ low-frequency modes. Since all networks share the same fourfold rotational symmetry axes, the constraint $U_{iu,jv} = U_{ij}\delta_{uv}$ still holds, and the band-structure is isotropic, i.e., the iso-frequency surface remains spherical. However, due to the large number of the internal degrees of freedom of the effective field, these modes are in general non-degenerate. The numerical simulation for the four-network example in Fig. 1(b), with $x_i$'s chosen to be 0, $5a/24$, $9a/24$, and $a/2$, indeed observes three non-degenerate low-frequency bands, in spite of the cubic symmetry of the structure. The fields for each of the three modes are strongly localized between two adjacent networks.

In the $N$-network structures, since all the low-frequency modes are linear, the density of electromagnetic states scales as $N-1$. Thus, the density of state can be greatly enhanced over very broad bandwidth, possibly extending from microwave to infrared, which can have important consequences for radiation and spontaneous emission control. In contrast, previous attempts to control spontaneous emission involve either broad-band suppression using photonic band gap [22], or narrow-band enhancement using microcavities [23]. It is also known that the electromagnetic fields between the metal regions are strongly enhanced. Thus, with nonlinear materials introduced between the metal regions, the nonlinear properties of these materials could provide a very interesting experimental system of studying the nonlinear properties of $N$-component fields, which is of considerable interest in modern field theory but until now has largely remained only as a theoretical novelty [24]. Finally, the three-network systems, with two low-frequency modes, might be designed to function as a transparent electrode that supports deep sub-wavelength propagating modes. Such electrode is important for light emitting diodes and photovoltaics applications.

*Case (3): Four-network FCC crystal.* The structure as shown in Fig. 1(c) is a four-network structure with a pyritohedral ($T_h$) symmetry. Each network [Inset in Fig. 4(a)] is identical in shape, and is made by translating $x$, $y$, and $z$-directional wires along the edges of a cube by $\pm b\hat{\mathbf{y}}$, $\pm b\hat{\mathbf{z}}$, and $\pm b\hat{\mathbf{x}}$, respectively, where $b = a/8$. These wires are then connected by a small cube at each site of a face-centered cubic (FCC) lattice. The basis element is topologically identical to Borromean rings [25]. This structure has an FCC lattice, and possesses four threefold rotational symmetries in diagonal directions and mirror symmetries in $x$, $y$, and $z$ directions. Based upon the symmetries, the **S** and **U** matrices become

$$\mathbf{S}=S_{12}\begin{pmatrix} -3 & 1 & 1 & 1 \\ 1 & -3 & 1 & 1 \\ 1 & 1 & -3 & 1 \\ 1 & 1 & 1 & -3 \end{pmatrix}, \mathbf{U}=\begin{pmatrix} \mathbf{U}_{11} & \mathbf{U}_{12} & \mathbf{U}_{13} & \mathbf{U}_{14} \\ \mathbf{U}_{12} & \mathbf{U}_{11} & \mathbf{U}_{14} & \mathbf{U}_{13} \\ \mathbf{U}_{13} & \mathbf{U}_{14} & \mathbf{U}_{11} & \mathbf{U}_{12} \\ \mathbf{U}_{14} & \mathbf{U}_{13} & \mathbf{U}_{12} & \mathbf{U}_{11} \end{pmatrix},$$

$$\mathbf{U}_{12}=\begin{pmatrix} U_{1x,2x} & 0 & 0 \\ 0 & U_{1y,2y} & 0 \\ 0 & 0 & U_{1z,2z} \end{pmatrix}, \mathbf{U}_{13}=\begin{pmatrix} U_{1z,2z} & 0 & 0 \\ 0 & U_{1x,2x} & 0 \\ 0 & 0 & U_{1y,2y} \end{pmatrix}, \quad (6)$$

$$\mathbf{U}_{14}=\begin{pmatrix} U_{1y,2y} & 0 & 0 \\ 0 & U_{1z,2z} & 0 \\ 0 & 0 & U_{1x,2x} \end{pmatrix}, \mathbf{U}_{11}=-(\mathbf{U}_{12}+\mathbf{U}_{13}+\mathbf{U}_{14}).$$

The eigenmodes now depend on the direction of the k-vector. The iso-frequency surfaces in the k-space for the three modes are all ellipsoids that are rotated versions of one another, having the major axis in *x*, *y*, and *z* directions, respectively [Fig. 4(b)]. These ellipsoids intersect at the [111] wavevector direction where all three modes are degenerate. Along the [111] direction the field profiles of the three modes are of an identical shape and rotated versions of one another.

The frequencies of the modes are determined by the three parameters $S_{12}U_{1x,2x}$, $S_{12}U_{1y,2y}$, and $S_{12}U_{1z,2z}$ in Eq. (6). In comparing theory to the FDTD simulations, we obtain these parameters by calculating the frequencies of the three modes at a single k-point where the modes are non-degenerate, e.g., $(0, 0, 0.04\pi/a)$. The full dispersion relation along all directions is then obtained using Eq. (5), and compared to FDTD simulations. The excellent agreements between the theory and simulations [Fig. 4(c)] provide a strong validation of our effective field approach.

The interlocking network topology here may be amenable to fabrication techniques such as self-assembly using block-copolymers [26]. While the simulation use perfect electric conductor, it should be reasonable even in the infrared region where plasmon responses are weak.

The eigenmodes of these structures at a given k-point have zero average electric field on the plane normal to $\vec{k}$. Therefore, the structures in general have zero coupling with an external plane wave normally incident on the structure. Components defined inside such structure are therefore inherently immune from radiation losses. Alternative, by placing arrays of dipole antennae at the surface of the structure, where the arms of the antennae are connected to different networks, the coupling between external plane waves and the internal modes can reach 100% [27].

This work is supported in part by ARO (Grant No. DAAD-19-03-1-0227), AFOSR (Grant No. FA9550-04-1-0437), NSF (Grant No. ECS-0134607), and a Samsung Scholarship. S. F. acknowledges discussions with J. B. Pendry.

**Figures**

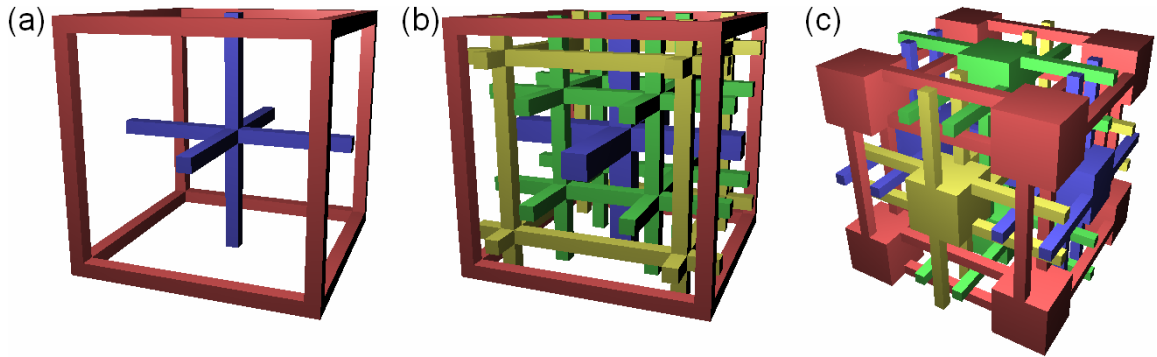

FIG. 1. Interlocking metallic-network structures. Structures in a single cubic unit cell are depicted. Each color represents a separate metallic network. (a) Two-network structure; (b) four-network cubic structure; (c) four-network FCC structure.

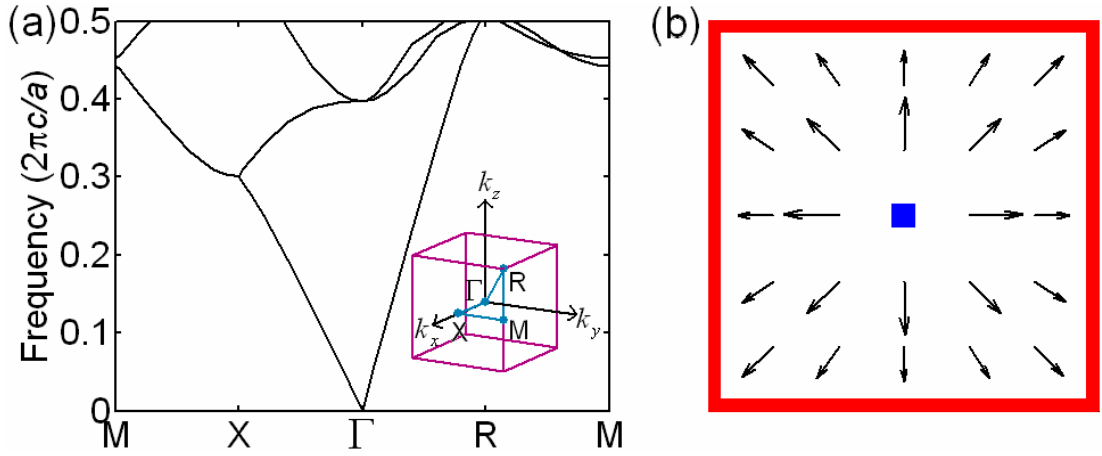

FIG. 2. (a) Band structure for the structure shown in Fig. 1(a). The inset shows the first Brillouin zone of a cubic lattice. (b) Electric field profile on the $z=0$ plane, at $\omega = 0.033(2\pi c/a)$ and $k = 0.05(2\pi/a)$, where $c$ is the speed of light in vacuum,

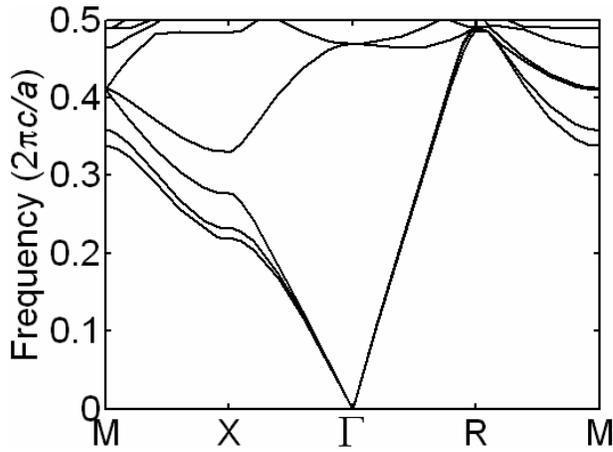

FIG. 3. Band structure of the four-network cubic system as shown in Fig. 1(b).

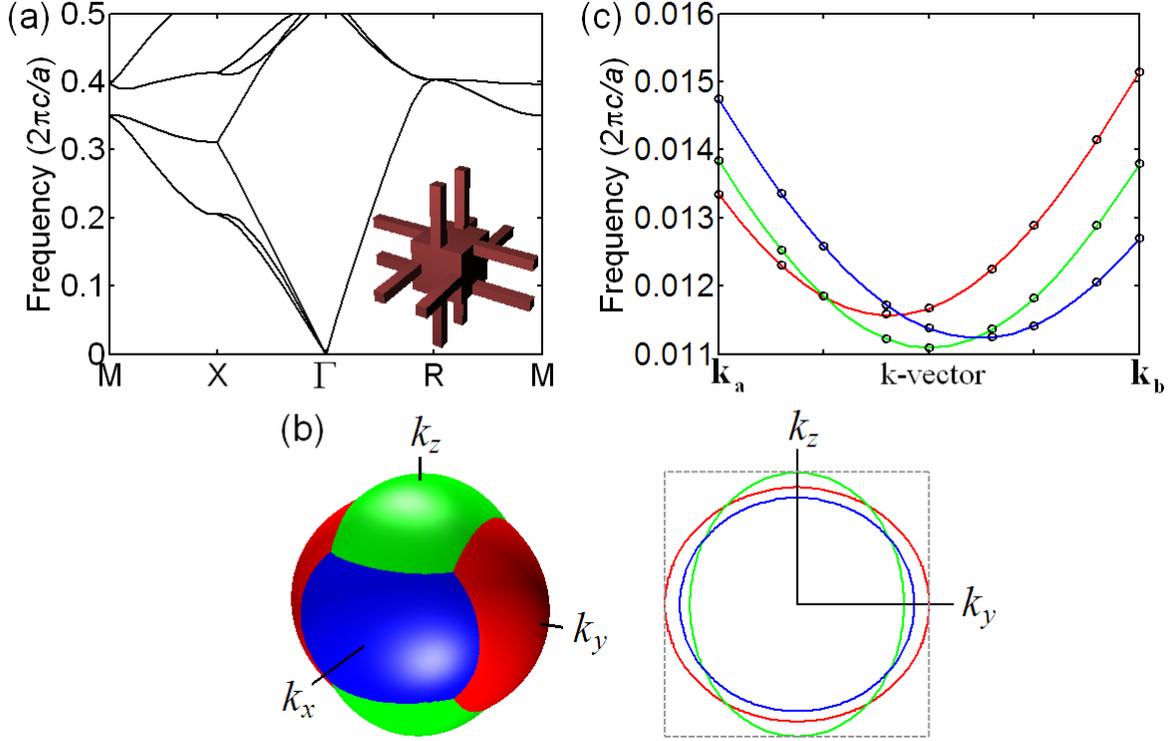

FIG. 4. Four-network FCC system: (a) band structure, (inset) the shape of one network, (b) iso-frequency surfaces and contours of three low frequency modes. The contour plot on the right is a $k_x=0$ slice. Gray dashed line represents a square that is tangential to some of contours. (c) Frequencies of modes for k-vectors along an $\mathbf{k_a}$–$\mathbf{k_b}$ line segment, where $\mathbf{k_a}$ = (0.0025, 0.016, 0.0175) × $2\pi/a$ and $\mathbf{k_b}$ = (0.0225, 0, 0.0075) × $2\pi/a$. The lines are theory predictions and the circles are FDTD results. These k-points are arbitrarily chosen. The agreement between theory and simulation is in fact excellent for all k-points.